\documentclass[conference]{IEEEtran}
\IEEEoverridecommandlockouts
\usepackage{cite}
\usepackage{amsmath,amssymb,amsfonts}
\usepackage{algorithmic}
\usepackage[pdftex]{graphicx}
\usepackage{textcomp}
\usepackage{xcolor}

\makeatletter
\def\input@path{{./tables/}}
\makeatother

\usepackage{lipsum,url,booktabs,multirow}
\graphicspath{{./figures/}}

\usepackage[caption=false,font=footnotesize]{subfig}

\newcommand{\tcell}[1]{\begin{tabular}[x]{@{}l@{}}#1\end{tabular}}

\makeatletter
\newcommand{\linebreakand}{%
  \end{@IEEEauthorhalign}
  \hfill\mbox{}\par
  \mbox{}\hfill\begin{@IEEEauthorhalign}
}
\makeatother

\def\BibTeX{{\rm B\kern-.05em{\sc i\kern-.025em b}\kern-.08em
    T\kern-.1667em\lower.7ex\hbox{E}\kern-.125emX}}
\begin{document}

\title{Comparing Generator Unavailability Models with\\Empirical Distributions from Open Energy Datasets\\
\thanks{This work was funded by the Supergen Energy Network (EP/S00078X/2) through the Climate-Energy Modelling for Assessing Resilience: Heat Decarbonisation and Northwest European Supergrid (CLEARHEADS) project.}
}

\author{
\IEEEauthorblockN{Matthew Deakin,\\David Greenwood}
\IEEEauthorblockA{\textit{School of Engineering} \\
\textit{Newcastle University}\\
Newcastle-upon-Tyne, UK \\
\{matthew.deakin, david.greenwood\}@ncl.ac.uk
}
  \and
\IEEEauthorblockN{~\vspace{-0.8em}\\David J. Brayshaw}
\IEEEauthorblockA{\textit{Meteorology Department} \\
\textit{University of Reading}\\
Reading, UK \\
d.j.brayshaw@reading.ac.uk}
%  \linebreakand % <------------- \and with a line-break
\and
\IEEEauthorblockN{~\vspace{-0.8em}\\Hannah Bloomfield}
\IEEEauthorblockA{\textit{School of Geographical Sciences} \\
\textit{University of Bristol}\\
Bristol, UK \\
hannah.bloomfield@bristol.ac.uk}
}

\maketitle

\begin{abstract}
The modelling of power station outages is an integral part of power system planning. In this work, models of the unavailability of the fleets of eight countries in Northwest Europe are constructed and subsequently compared against empirical distributions derived using data from the open-access ENTSO-e Transparency Platform. Summary statistics of non-sequential models highlight limitations with the empirical modelling, with very variable results across countries. Additionally, analysis of time sequential models suggests a clear need for fleet-specific analytic model parameters. Despite a number of challenges and ambiguities associated with the empirical distributions, it is suggested that a range of valuable qualitative and quantitative insights can be gained by comparing these two complementary approaches for modelling and understanding generator unavailabilities. \end{abstract}


\section{Introduction}
Power station unavailability modelling is an essential part of power system planning and operations. It is necessary for reserve scheduling to ensure second-by-second stability, hourly unit commitment, or resource adequacy planning several years into the future. Despite this reputation as a textbook problem, insufficient mitigation against supply-side vulnerabilities was found to have been a key contributing factor in the catastrophic Texas blackouts of February 2021 \cite{busby2021cascading}. Furthermore, nascent Digital Twins of energy systems, incorporating highly granular system-specific data, will need to provide a virtual replica of all parts of the energy system so that they can evaluate system security as a part of their core functionality \cite{ngeso2021virtual}.

It is well-known that different generator types (e.g., nuclear, gas) have different availabilities \cite{edwards2017assessing}. However, for accurate risk assessment, the overall unavailability of generator fleets also depends on characteristics of the fleet itself, such as the age and condition of units, as well as exogenous factors such as meteorological conditions \cite{murphy2020resource}. Databases containing a large number of generator outage reports are necessary for considering these issues, as there are typically hundreds of large generators in countries such as the UK or Germany.

One such database is provided by the European Network of Transmission System Operators for Electricity (ENTSO-e), who have been providing the Transparency Platform continuously since 2015 \cite{entsoe2021transparency}. Compared to other data that are presented on the platform, generator unavailability data is more complex \cite{hirth2018entso}, from the point of view of the volume of the data, but also its interpretability (e.g., the difference between the self-reported `forced' and `planned' outages). Indeed, previous works that make extensive use of data from the platform stop short of working with the outage data stream directly \cite{rafiee2021data}.

Given the challenges with working with this type of data, unavailability modelling and analysis using empirical data directly is not common. In \cite{murphy2020resource}, the authors consider the implications of temperature dependence on the PJM system in the USA, finding that temperature dependence of generators significantly increases the capacity that must be procured to achieve a given risk standard. A similar conclusion is found in \cite{golub2020does}, again focusing on high temperatures and drought in the German system. Further use-cases for consider generator models beyond system adequacy are considered, for example, in \cite{haluvzan2020performance}, where the authors use outage data from the ENTSO-e Transparency Platform for improving forecasting. In \cite{wels2019quality}, the author explores the challenges and possible solutions to mixed data quality in the ENTSO-e unavailability data, considering the Dutch and German systems. The authors of \cite{lorencin2016evaluating} consider Bayesian priors as an alternative and more accurate method of power system outages by modelling plants with missing or incomplete unavailability data with plants that are similar. 

To our knowledge, there are no papers which compare analytic fleet unavailability models with empirical data across a wide range of countries. This is an important and timely gap--such unavailability modelling across countries is central to the administration of contemporary capacity markets (such as the ca. £700m GB capacity market \cite{ngeso2021modelling}). Furthermore, as databases such as the Transparency Platform become increasingly accessible, it becomes even more crucial for the limitations of given datasets to be explored in detail.

The contribution of this work is to address this gap by comparing analytic, aggregate unavailability models for a fleet of generators against equivalent empirical models derived from an open energy database (the ENTSO-e Transparency Platform). Both time collapsed (non-sequential) and time sequential models are considered for a range of countries, and limitations of the empirical data are described.

This paper outlines analytic unavailability models (Section~\ref{s:modelling}); the approach used to download the empirical unavailability data (Section~\ref{s:processing}) and then compares these two approaches (Section~\ref{s:results}) before offering conclusions (Section~\ref{s:conclusions}), to highlight similarities and differences between these complementary approaches of unavailability modelling.

\section{Analytic Generator Unavailability Models}\label{s:modelling}

This work considers two main classes of model: time collapsed (or `non-sequential') models, which study the distribution of plant outages in the long run, and time sequential models, which consider the hour-by-hour time series of generator unavailabilities. In this section, we first highlight models that are most commonly used to model unavailability of an individual generator, before describing how these can then be combined to create the system-wide unavailability.

\subsection{Time collapsed outage modelling}\label{ss:modelling_timecol}

Time collapsed models are based on the approximation that a unit's availability is based a model which can be determined by an availability parameter, $A$, which determines the likelihood of a generator being available at a given time instant. An individual production unit $U$ with size $U_{0}$ (in~MW) and availability $ A $ can then modelled as a Bernoulli random variable,
\begin{equation}\label{e:t_coll_bernoulli}
\dfrac{U}{U_{0}} \sim \mathrm{Bernoulli}(A)\,.
\end{equation}
It is well-known that the probability density function of the sum of independent random variables can be determined by the convolution of each of the PDFs of each random variable. Therefore, under the assumption of independent outages, the PDF of total system outages can be determined by calculating the convolution of the PDFs of each individual production unit \cite[Ch. 3]{billinton1996reliability}.

\subsection{Time Sequential outage modelling}\label{ss:modelling_timeseq}

Time sequential outage modelling explicitly considers how a unit's availability changes in time. The most parsimonious model of the outages of a given generator is via a two-state Markov model: a production unit is modelled as being either in or out of service. In each of these states, there is a transition probability that describes the likelihood of moving to the other state. 

These transition probability can be specified directly, or they can be determined by other parameters. For example, it is common to describe the outage states of generator models via an availability parameter $ A $ and Mean Time To Repair, MTTR (in hours). The repair rate $ \mu $ and failure rate $ \lambda $ can be determined from these parameters \cite[Ch. 9]{billinton1992reliability} as
\begin{equation}\label{e:trn_prob}
\mu = \dfrac{1}{\mathrm{MTTR}}\,, \qquad \lambda = \mu\left (\dfrac{1}{A} - 1\right ) \,.
\end{equation}

\subsection{Generator Physical Availability Models for Northwest Europe}\label{ss:modelling_fleets}

To build total outage rates for the generator fleet of individual balancing zones across Northwest Europe, the total installed capacity for each fuel type is first found from the `Installed Production Data' from ENTSO-e. Individual Production and Generation units sizes for all countries are pooled to create an empirical distribution of reasonable generator sizes for each fuel type. From this, a representative fleet of generators is then created by allocating these generators to each country and fuel type. Generators of a given fuel type are assumed to all have the same availability and MTTR, with the values used summarised in in Table~\ref{t:tbl_fuel_mttrs} (MTTRs are from \cite{edwards2017assessing} and availabilities are from \cite{ngeso2020ecr}).

\begin{table}
% Generated using basicTblSgn.
\centering
\caption{Availability $A$ and Mean Time To Repair (MTTR) by fuel type.}\label{t:tbl_fuel_mttrs}
\begin{tabular}{lllllll}
\toprule
Fuel & Avlbty., $A$ & \tcell{MTTR\\(hrs)} &  & Fuel & Avlbty., $A$ & \tcell{MTTR\\(hrs)} \\
\midrule
Biomass & 0.86 & 40 &  & Coal & 0.86 & 40 \\
CCGT & 0.90 & 50 &  & Oil & 0.91 & 50 \\
Hydro & 0.90 & 20 &  & Nuclear & 0.81 & 150 \\
CHP & 0.90 & 50 &  & Waste & 0.86 & 40 \\
\bottomrule
\end{tabular}

\label{t:tbl_fuel_mttrs}
\end{table}

For example: if the UK in 2020 had 10 GW of nuclear power, and across Europe nuclear production units are only 1 GW or 2 GW units with equal probability, then one possible nuclear generation fleet for the UK could be three 2 GW units and four 1 GW units. Each of these would have the same availability and mean time to repair (Table~\ref{t:tbl_fuel_mttrs}).

\section{Data Processing for Empirical Unavailability Distributions}\label{s:processing}

In this section, we describe how the outage data for individual units are accessed from the ENTSO-e platform and subsequently processed to estimate the aggregate unavailabilty of the whole fleet.

The Transparency Platform reports Planned and Forced outages, thereby covering both `planned' and `actual' unavailability as required by the relevant regulations \cite{eu2013transparency}. The regulation specifically states that plant of capacity greater than 200~MW must report changes of more than 100~MW in availability, both for aggregate generation units or individual production units. 

In practise, changes in availability much smaller than 100~MW are reported extensively by many units (see Section \ref{ss:results_example_outages}). Users of this data should therefore be cognizant that there could be small outages which are not reported. 

\subsection{Data Access Approach}\label{ss:processing_defining_outages}

The data are downloaded using the Transparency Platform API, allowing for automated download, with the data downloaded as XML files for individual days. This means that some reports are duplicated (particularly where outages last many months), but it means that the data can be inspected manually if required. Our focus is on dispatchable generators, and so unavailability of renewable technologies (solar, onshore wind, offshore wind, and run-of-river hydro) is neglected. Furthermore, reports with a `withdrawn' flag are also ignored.

\begin{figure}\centering
\includegraphics[width=0.44\textwidth]{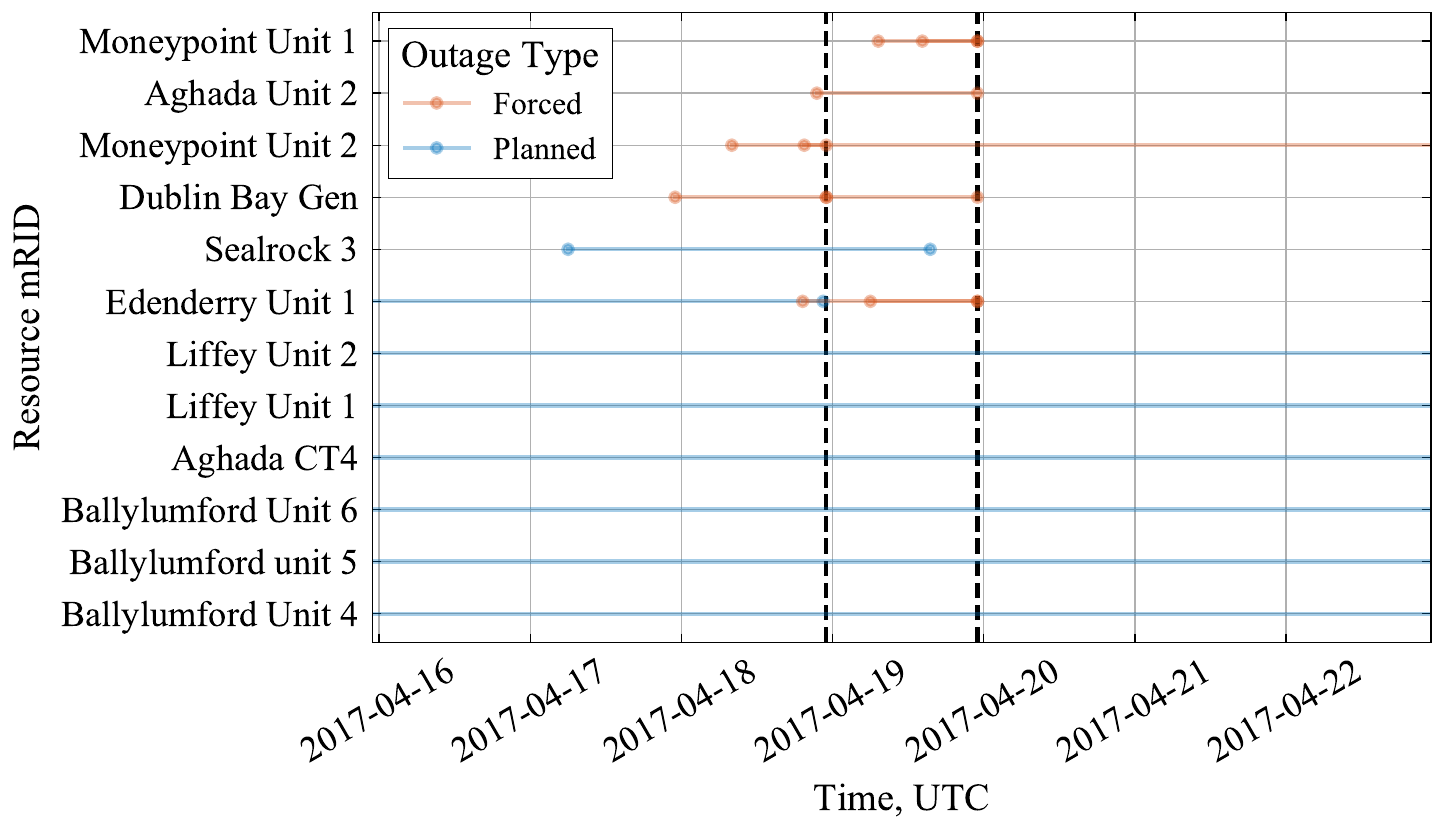}
\caption{The outage reports for the IE zone for 19/4/17. The solid dots denote the start and end point of a single outage report for a production unit.}\label{f:fig_entsoeout}
\end{figure}

A representation of the reports downloaded for one day of the IE zone is shown in Fig.~\ref{f:fig_entsoeout}. It can be observed that, for some generators, there are multiple overlapping reports for the given day, with some generators having both forced and planned outages. It can also be seen that some of the generators have long-term outages, with their planned outage extending beyond the week window shown here.

\subsubsection{Defining Hourly Outages}

Outage reports are given at a temporal resolution of one minute, and so over the course of an hour, the unavailability may change multiple times. For this work, the `outage' for a given hour is the mean outage over the given time period--i.e., if the `instantaneous' (minute-by-minute) outage is denoted by $o$, the hourly outage $ O(\tau) $ considered over some time period $ \tau,\, \tau+\delta_{\tau} $ is
\begin{equation}
O(\tau) = \dfrac{1}{\delta_{\tau}}  \int_{\tau}^{\tau+ \delta_{\tau}} o(t)\, d 
t\,.
\end{equation}
For example, if a production unit's availability is reduced by 50~MW for 12 minutes and 200~MW for 48 minutes, then the corresponding hour's outage would be calculated to be 170~MW.

In some cases, a given unit will have multiple outage reports which are contradictory at a particular time. In these cases, the conflicting outage reports need to be reconciled. The approach in this work is to combine these multiple conflicting reports $o_{1},\,\ldots\,,\,o_{n}$ and then calculate the minimum, maximum, and mean outage that could occur given all of those reports, as
\begin{align}
O_{\min}(\tau) &= \dfrac{1}{\delta_{\tau}} \int_{\tau}^{\tau+ \delta_{\tau}} 
\min _{i}\left \{ o_{i}(t) \right \}\, d t\,, \label{e:recon_min}\\
O_{\max}(\tau) &= \dfrac{1}{\delta_{\tau}} \int_{\tau}^{\tau+ \delta_{\tau}} 
\max _{i}\left \{ o_{i}(t) \right \}\, d t\,, \label{e:recon_max}\\
O(\tau) &= \dfrac{O_{\min}(\tau) + O_{\max}(\tau) }{2}\,, \label{e:outage_def}
\end{align}
where the $ \min $ and $ \max $ operators calculate the maximum and minimum value over all of the outage reports at each incremental time period. 

A further common inconsistency that is found is that, in some cases, the reported reduction in capacity is greater than the total size of the generator itself (e.g., a reduction in capacity of 1500~MW may be reported for a generator whose size is only 750~MW). For the purposes of this work, outage reports for which the reported outage is 33\% greater than the production unit size are ignored (outages slightly above the size of the plant are permitted, as it allows for small inaccuracies in the reporting of unit maximum ratings which may exist).

Forced and planned outages are recorded separately, as well as the total outage level. The latter is determined by calculating the outages irrespective of the type of outage flag. Note that the total outages are always less than or equal to the sum of the forced and planned outages, with conflicting forced and planned outages reconciled to calculate the total outage through \eqref{e:outage_def} (i.e., neglecting their state as forced or planned).

\subsection{Evaluating Data Quality}

Clearly, as highlighted in the previous section, there are a range of reasons as to why the unavailabilities of the aggregated fleet of individual countries may be inaccurate. In the first instance, only shortfalls of more than 100~MW are mandatory to report (although, in many cases shortfalls are reported at a much higher resolution than this). There is also potential for human error in the reports that are submitted to the Transparency Platform, and mothballed or closed plant are at times listed as being on an extended `outage'. Furthermore, there may be commercial reasons for generators to attempt to obfuscate the true outage state of their plant (e.g., to give traders within the same organisation as much of a competitive advantage as possible).

One of the uncertainties which can be considered explicitly is the effect that the reconciliation of conflicting reports \eqref{e:outage_def} could have. In particular, we calculate the relative mean absolute reconciliation error $ \epsilon $ over the full time period, i.e.,
\begin{equation}\label{e:recon_err}
\epsilon = \dfrac{\| O(\tau) - O_{\min}(\tau) \| _{1} }{ \| O(\tau) \|_{1} } \;.
\end{equation}
This relative reconciliation error gives a measure as to the amount that the outages could be incorrect given these conflicting outage reports. In other words, by comparing the reconciliation error $\epsilon$ with the mean and interquartile range (IQR) of the total outage distributions, we can confirm if the reconciliation is likely to have made a significant impact on the final distribution.

\section{Results}\label{s:results}

In this section, we look to consider the differences between the empirical and modelled system outages. In total, data was downloaded for five winters and four summers for ten regions. The number of reports per day varies between countries--the smaller Irish system regularly has fewer than ten reports per day, whilst the German and French systems regularly have more than two hundred.

In this work, we focus on modelling outages during the winter period, as this is the season which has peak demands in Northwestern Europe. For the purposes of this, we consider the winter season to consist of twenty weeks following the first Sunday of November, excluding two low-demand weeks around Christmas \cite{deakin2021impacts}.

In this section, we first present the unavailability profiles of a number of generators to demonstrate clearly some of the challenges of working with the outage data directly and to consider how this empirical data (Section \ref{s:processing}) differs qualitatively from the presented analytic models (Section \ref{s:modelling}). The total fleet unavailabilities are then considered for eight Northwest European countries, and the properties of these empirical models compared against both time sequential and time collapsed models. The seasonality of the empirical data is then explored, followed by a short discussion to consider implications of the results for future unavailability modelling.

The aggregated unavailability for each country (and code used to derive these unavailabilities) is available from \cite{deakin2022aggregated}.

\subsection{Example Data}\label{ss:results_example_outages}

\begin{figure}\centering
\subfloat[Example outage reports, Jan-Mar 2019.]{\includegraphics[width=0.44\textwidth]{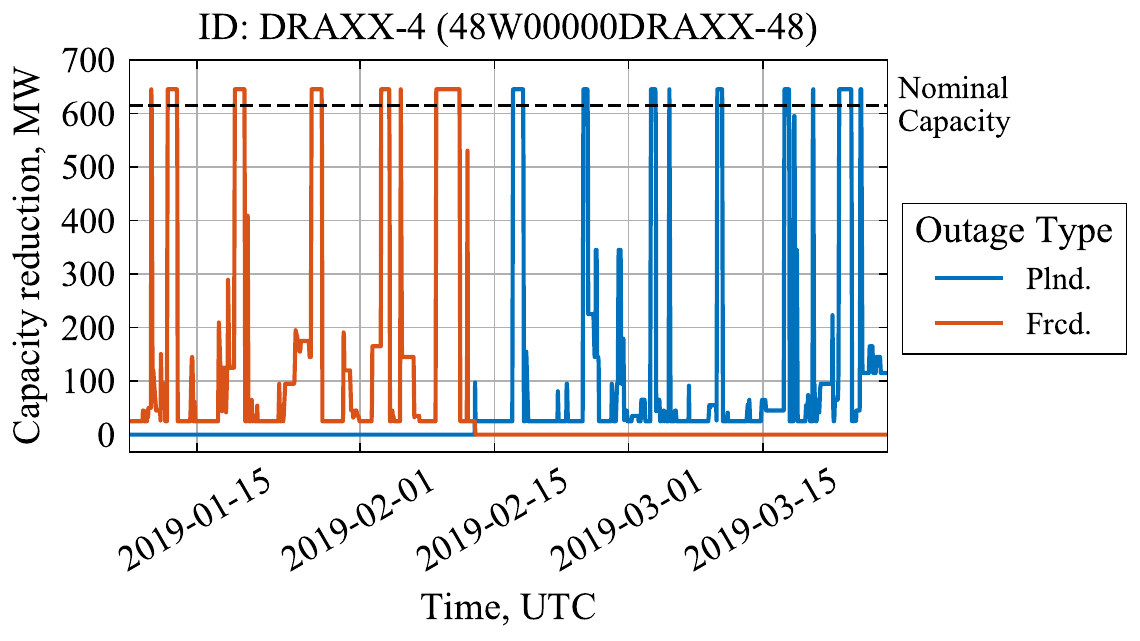}\label{ff:plt_xmpl_4}}
~\\
\subfloat[Example outage reports, Jan 2021.]{\includegraphics[width=0.44\textwidth]{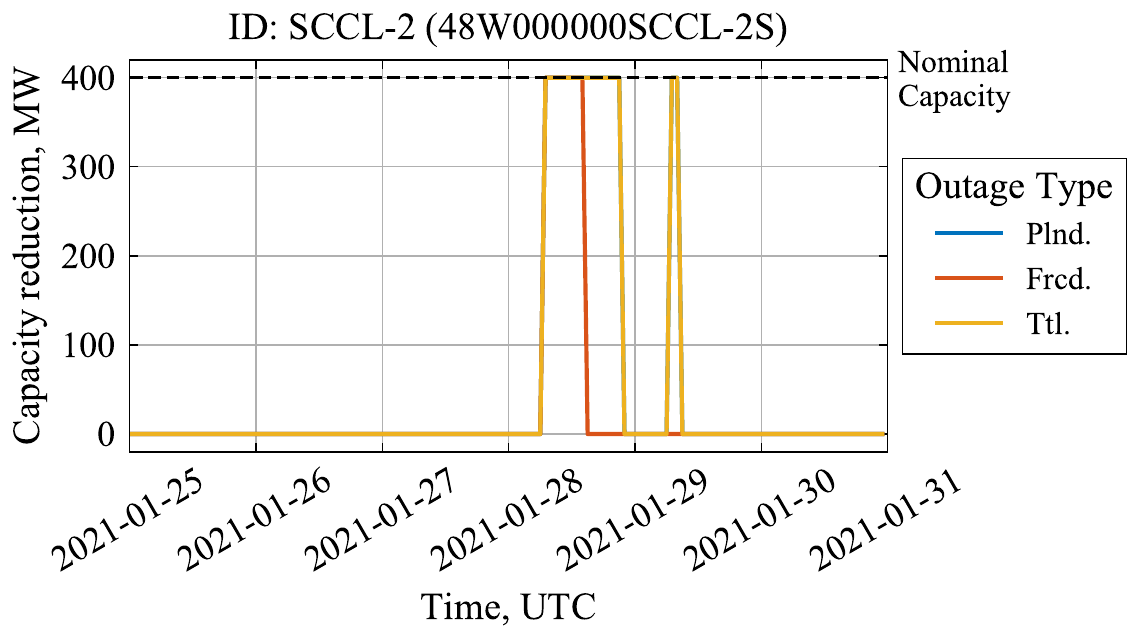}\label{ff:plt_xmpl_5}}
~\\
\subfloat[Example outage reports, Oct. 2016.]{\includegraphics[width=0.44\textwidth]{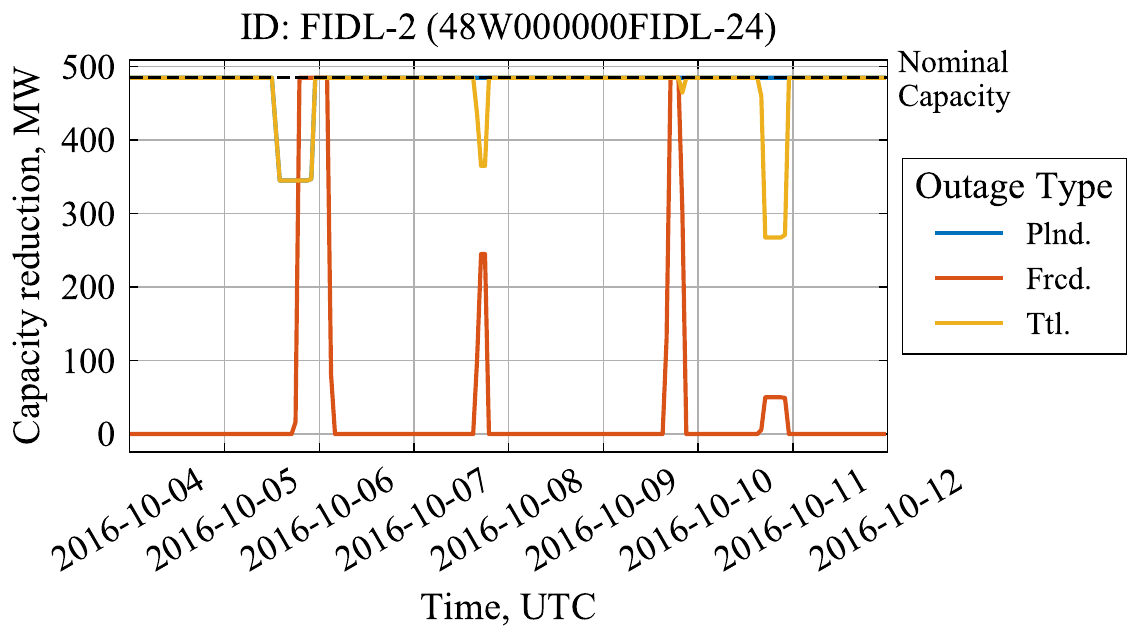}\label{ff:plt_xmpl_3}}
\caption{Generator unavailability for three units in the GB system.}
\label{f:plt_xmpls}
\end{figure}

First, we consider the unavailability reports for a number of individual periods. Fig.~\ref{f:plt_xmpls} plots the planned, forced and total outages for three production units.

Fig.~\ref{ff:plt_xmpl_4} highlights how the distinction between forced (Frcd.) and planned (Plnd.) outages is not clear-cut. For example, it can be seen that, around mid February, the forced outages from earlier in the winter are instead classified as planned outages. By eye, there appears to be a clear pattern in the outages, and so the reason for the change in classification is not clear. Note also how small capacity reductions (less than 100~MW) are reported regularly.

In Fig.~\ref{ff:plt_xmpl_5}, it can be why the Total (Ttl.) fleet outages are not necessarily a linear sum of Forced and Planned outages (as mentioned in Section \ref{ss:processing_defining_outages}). In this case, as the nominal capacity of the unit is 400~MW, on the day that there is 400~MW of reduction in both planned and forced outages, the approach outlined in Section~\ref{ss:processing_defining_outages} returns a total outage of only 400~MW (rather than 800~MW). In other words, it is assumed that these are for the same incident.
 
Finally, Fig.~\ref{ff:plt_xmpl_3} shows a more complex example: for this plant, there is an active report that states that unit has a planned outage from 1st October 2016 to 30th September 2017, with capacity of the unit reduced by 485~MW. However, during this time period, there are further active outage reports which clearly contradict this first report (as shown in the figure), and tending to lead the mean outage to drop when these outages are not at the nominal capacity of the unit (due to the approach mentioned in Section \ref{ss:processing_defining_outages}). It is beyond the scope of this work to attempt to comprehensively identify and correct apparent errors such as these, although more systematic ways of considering these types of errors could improve accuracy further.

\subsection{Time Collapsed Model Comparison}

\begin{figure}\centering
\includegraphics[width=0.4\textwidth]{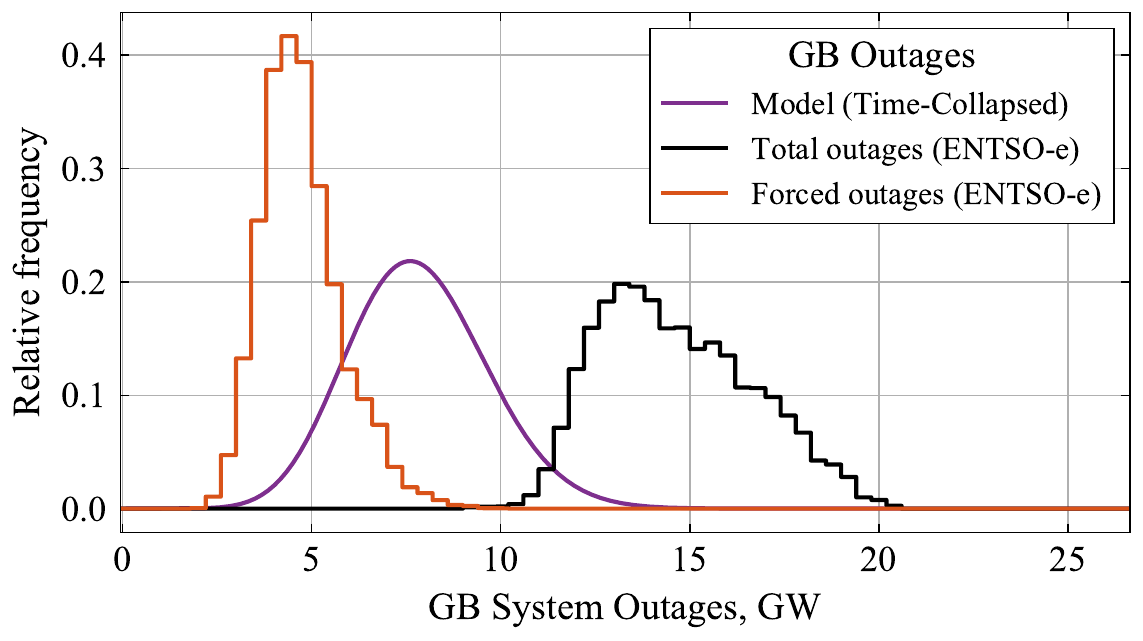}
\caption{Histograms of the Total and Forced outages for the GB system from the ENTSO-e platform for five winters (16/17 to 20/21), compared against the PDF of the analytical Time-Collapsed model of the same system.}\label{f:pltCompareOutages_GB}
\end{figure}

The time collapsed model of total outages for the fleet can be described by the PDF of the total generator unavailabilities. In Fig.~\ref{f:pltCompareOutages_GB}, histograms of the total and forced outage rates in for five winter seasons are compared against the analytic model based on \eqref{e:t_coll_bernoulli}. It can be observed that in this case, the spread of the modelled outages falls roughly between the forced and total outages. At first glance, this appears to show that the analytic model and empirical model show some links, even if there are clear discrepancies.

However, once a full range of countries are considered, it is quite clear that in many cases the empirical data are far from being a good representation of the analytic model--the mean and IQR of the analytic model and empirical data are given in Table~\ref{t:tblSumStats}. For example, countries such as Belgium and Norway show means and IQRs that are many times smaller than the modelled outage rates. This suggests a clear discrepancy between the empirical and analytic models; further investigations are clearly required to provide an accurate picture of generator unavailability in those systems. Note that although the reconciliation error \eqref{e:recon_err} is non-trivial, it clearly is relatively small as compared to the bulk errors observed and so does not explain those discrepancies.

\begin{table}
% Generated using basicTblSgn.
\centering
\caption{Mean and Interquartile Range (IQR) of Forced (Frcd.), Total (Tot.), and Modelled (Mld.) total system outages for nine countries, alongside the maximum reconciliation error \eqref{e:recon_err}.}\label{t:tblSumStats}
\begin{tabular}{rlllllll}
\toprule
\multirow{2}*{\vspace{-0.4em} \begin{tabular}[x]{@{}r@{}}Ctry.\\Code\end{tabular} } & \multicolumn{3}{c}{Mean, GW} & \multicolumn{3}{c}{IQR, GW} & \multirow{2}*{\vspace{-0.4em} \begin{tabular}[x]{@{}l@{}}Max Recon.\\Error $\epsilon $, \%\end{tabular}}\\
            \cmidrule(l{0.6em}r{0.9em}){2-4} \cmidrule(l{0.6em}r{0.9em}){5-7} 
        & Frcd. & Ttl. & Mdl. & Frcd. & Ttl. & Mdl. \\\midrule
GB & 4.7 & 14.7 & 7.8 & 1.3 & 3.0 & 2.5 & 0.7\% \\
BE & 0.1 & 0.2 & 2.2 & 0.2 & 0.4 & 1.4 & 2.4\% \\
DE & 2.8 & 11.4 & 14.2 & 1.5 & 3.5 & 3.3 & 1.6\% \\
DK & 0.5 & 0.8 & 1.1 & 0.5 & 0.5 & 0.8 & 3.0\% \\
ES & 5.5 & 6.3 & 8.1 & 3.7 & 3.6 & 2.2 & 0.1\% \\
FR & 2.7 & 16.6 & 15.9 & 2.3 & 7.8 & 4.7 & 0.1\% \\
IE & 0.1 & 0.6 & 1.0 & 0.3 & 0.5 & 0.8 & 1.6\% \\
NL & 0.7 & 3.5 & 2.7 & 1.1 & 1.4 & 1.3 & 0.1\% \\
NO & 0.2 & 0.7 & 3.0 & 0.3 & 0.6 & 1.0 & 0.0\% \\
\bottomrule
\end{tabular}

\label{t:tblSumStats}
\end{table}

\begin{figure}\centering
\includegraphics[width=0.44\textwidth]{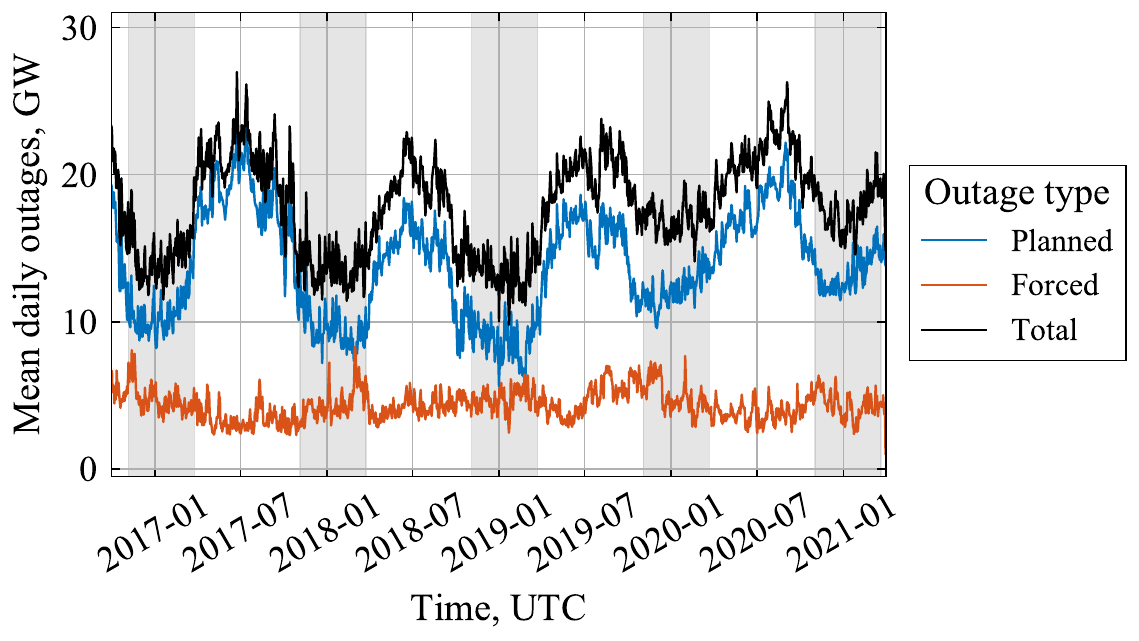}
\caption{Forced outages in the GB system are much less seasonal than Planned outages, and make up less than half of all outages in the GB system.}\label{f:pltTtlFcdTs}
\end{figure}

In any case, it is interesting to note that there are marked seasonal characteristics in the outage rates. For the GB system the Forced outages are, to a large extent, constant, whilst planned outages fluctuate on an annual basis, with a clear minimum during the winter months (Fig.~\ref{f:pltTtlFcdTs}).

\begin{figure}\centering
\subfloat[Mean weekly outages]{\includegraphics[width=0.23\textwidth]{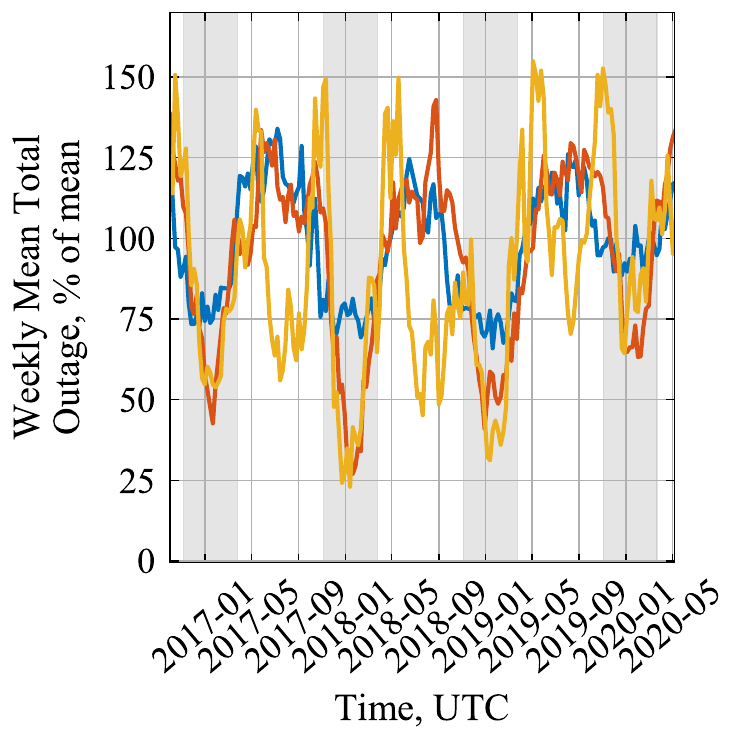}\label{ff:pltWeeklyCountry}}
~\subfloat[Mean weekly demand]{\includegraphics[width=0.26\textwidth]{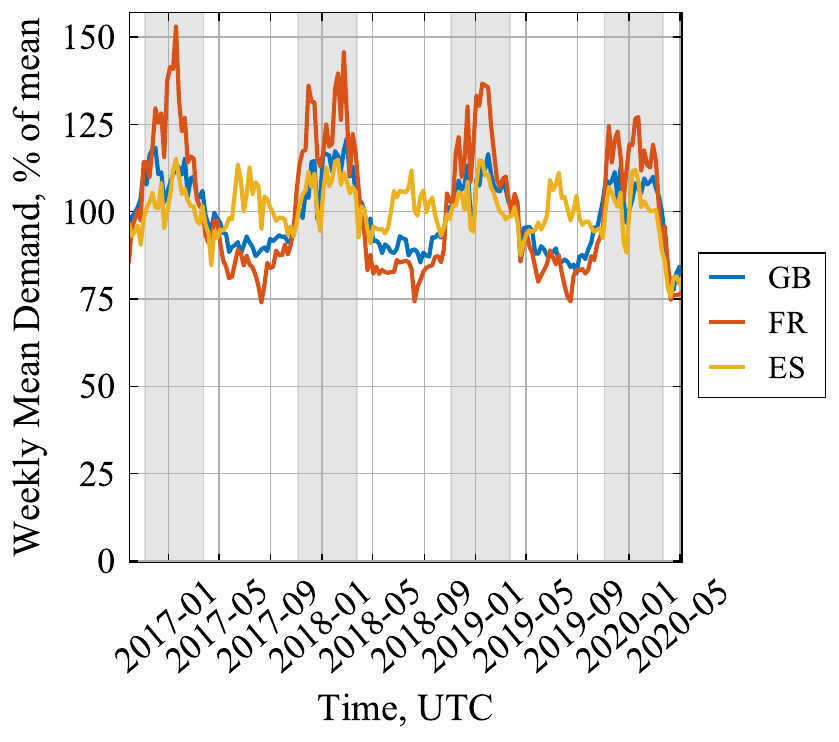}\label{ff:pltWeeklyDemand}}
\caption{The seasonality of reported outages varies from system to system, with a clear link between total outages (a) and the demand (b).}
\label{f:pltWeekly}
\end{figure}

Note that, whilst some seasonality appears to be universal, the extent and period of this seasonality varies considerably. The normalised mean weekly plant Total outages for GB, France and Spain are plotted in Fig.~\ref{f:pltWeekly} alongside the normalised seasonal demands. Fig.~\ref{ff:pltWeeklyCountry} shows how the GB system has less seasonality than either the Spanish or French systems, and that the Spanish system has a biannual cycle. This biannual cycle could be explained as being due to generators ensuring maintenance does not occur during demand peaks, of which there are two in the year in Spain, as compared to France or GB which only peak in the winter (Fig.~\ref{ff:pltWeeklyDemand}).

\subsection{Time Sequential Modelling}

Time sequential models are much more complex than non-sequential time collapsed models: a complete characterization of the analytic and empirical models is beyond the scope of this work. Here, we compare the data for GB visually, and calculate the signal autocorrelation across a range of countries.

\begin{figure}\centering
\subfloat[Empirical data]{\includegraphics[width=0.25\textwidth]{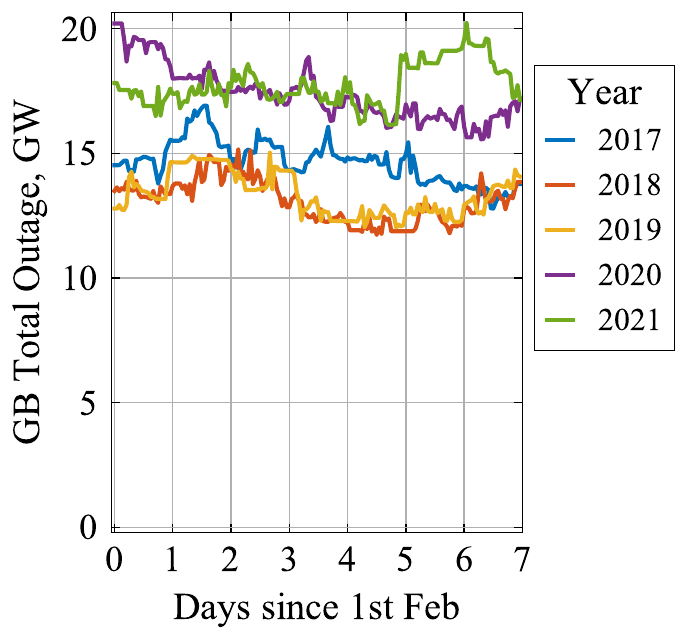}\label{ff:pltOneWeekXmpl}}
~~~
\subfloat[Model outputs]{\includegraphics[width=0.22\textwidth]{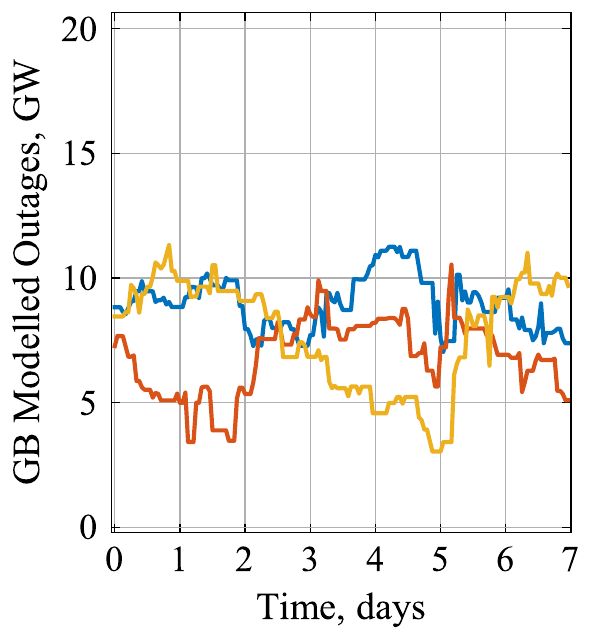}\label{ff:pltTsXmplRuns}}
\caption{The time series of total outages (Fig.~\ref{ff:pltOneWeekXmpl}) and modelled outages (Fig.~\ref{ff:pltTsXmplRuns}) show that, despite the spread of the model outputs being smaller than that of the Total Outages (Table~\ref{t:tblSumStats}), the hour-by-hour changes in model outputs are much greater than the changes from the empirical model.}
\label{f:pltTsComparison}
\end{figure}

Firstly, the Total outages for the days following Feb. 1st from 2017 to 2021 are compared against three draws of the two-state time sequential model (Section \ref{ss:modelling_timeseq}) in Fig.~\ref{f:pltTsComparison}. Clearly, the mean of the analytic model is lower than empirical data; this is not unexpected given Fig.~\ref{f:pltCompareOutages_GB}. However, what is also clear is that, despite the spread of the Total outages being greater than that of the model (as determined using the IQR in Table~\ref{t:tblSumStats}), the hour-by-hour \textit{changes} in the analytic model outputs appear to be greater than those of the reported Total outages from the GB system. Future work could systematically study these changes where they are critical (e.g., for storage scheduling during periods with tight margin).

Finally, in Table~\ref{t:tbl_acf}, the autocorrelation is tabulated for a variety of time lags for five countries for which the empirical data and analytic models appear reasonably consistent (considering the statistics in Table \ref{t:tblSumStats}). As the time series for the five winters are not contiguous, the mean of the autocorrelation for five winters (and five simulated winters) is calculated and reported in this table.

It is clear that for GB the autocorrelation is relatively well-calibrated based on the model data (despite the aforementioned disparity between the changes in the data). However, it can be seen that the unavailability data for other countries (particularly France and Spain) show much higher autocorrelation than the modelled data, particularly at time periods longer than one hour. This appears to suggest that the time sequential modelling for those countries is not accurate for longer time periods. Depending on the purposes of the modelling being carried out, this suggests that the Markov model parameters need tuning (or may even require a multi-state model), or that alternative black-box modelling approaches may be more appropriate (e.g., the use of Box-Jenkins type models).

\begin{table}
% Generated using basicTblSgn.
\centering
\caption{Comparing the mean of the autocorrelation of Total outages in Western European counries for five winters (16/17 - 20/21) against models, considering lags from one hour to one week.}\label{t:tbl_acf}
\begin{tabular}{rllllllll}
\toprule
\multirow{3}*{\vspace{-1em} \begin{tabular}[x]{@{}r@{}}Ctry.\\Code\end{tabular} } & \multicolumn{8}{c}{Autocorrelation at given lag} \\
        \cmidrule(l{0.6em}r{0.9em}){2-9}
        & \multicolumn{2}{c}{1 hr} & \multicolumn{2}{c}{6 hrs} & \multicolumn{2}{c}{1 day} & \multicolumn{2}{c}{1 week}\\
        \cmidrule(l{0.6em}r{0.9em}){2-3} \cmidrule(l{0.6em}r{0.9em}){4-5} \cmidrule(l{0.6em}r{0.9em}){6-7} \cmidrule(l{0.6em}r{0.9em}){8-9} 
    & Data & Mdl. & Data & Mdl. & Data & Mdl. & Data & Mdl.\\\midrule
GB & 0.97 & 0.97 & 0.86 & 0.85 & 0.62 & 0.53 & 0.27 & -0.01 \\
FR & 0.99 & 0.98 & 0.97 & 0.87 & 0.89 & 0.62 & 0.69 & 0.08 \\
DE & 0.99 & 0.97 & 0.92 & 0.81 & 0.72 & 0.45 & 0.35 & -0.04 \\
ES & 0.99 & 0.97 & 0.97 & 0.82 & 0.93 & 0.47 & 0.70 & -0.01 \\
NL & 0.98 & 0.97 & 0.91 & 0.85 & 0.77 & 0.49 & 0.26 & -0.05 \\
\bottomrule
\end{tabular}

\label{t:tbl_acf}
\end{table}

\section{Conclusions}\label{s:conclusions}

Accurate models of generator availability will become even more critical to ensure security of supply in energy systems based on a combination of renewable energy technologies and dispatchable peaker plant. This paper has considered how analytic generator unavailability models compare against empirical distributions derived from data sourced from the ENTSO-e Transparency Platform. Nine European systems were considered, centred on Northwestern Europe, with five winters studied based on tens of thousands of individual generator outage reports. 

The ex-post analysis of power system variables linked closely to generator unavailability, such as system margin, are likely to remain elusive without a step-change in data quality. For example, validating the representativeness of non-sequential analytic models using empirical data for countries such as Belgium and Norway does not appear feasible. More complex time sequential models appear to show quite different properties than the empirical time series they are intended to represent, considering attributes such as autocorrelation. 

Despite these complexities, it is concluded that empirical generator unavailability data brings a range of insights to the nature and workings of modern power and energy systems, and should feed into mathematical generation unavailability models. Future work could consider further data cleansing techniques, incorporating alternative sources of unavailability data (e.g., from system operator), and alternative time sequential modelling approaches.

\bibliographystyle{IEEEtran}
\bibliography{./outage_bib}{}

\end{document}